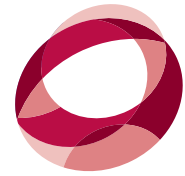

# Evolving Methods for Evaluating and Disseminating Computing Research

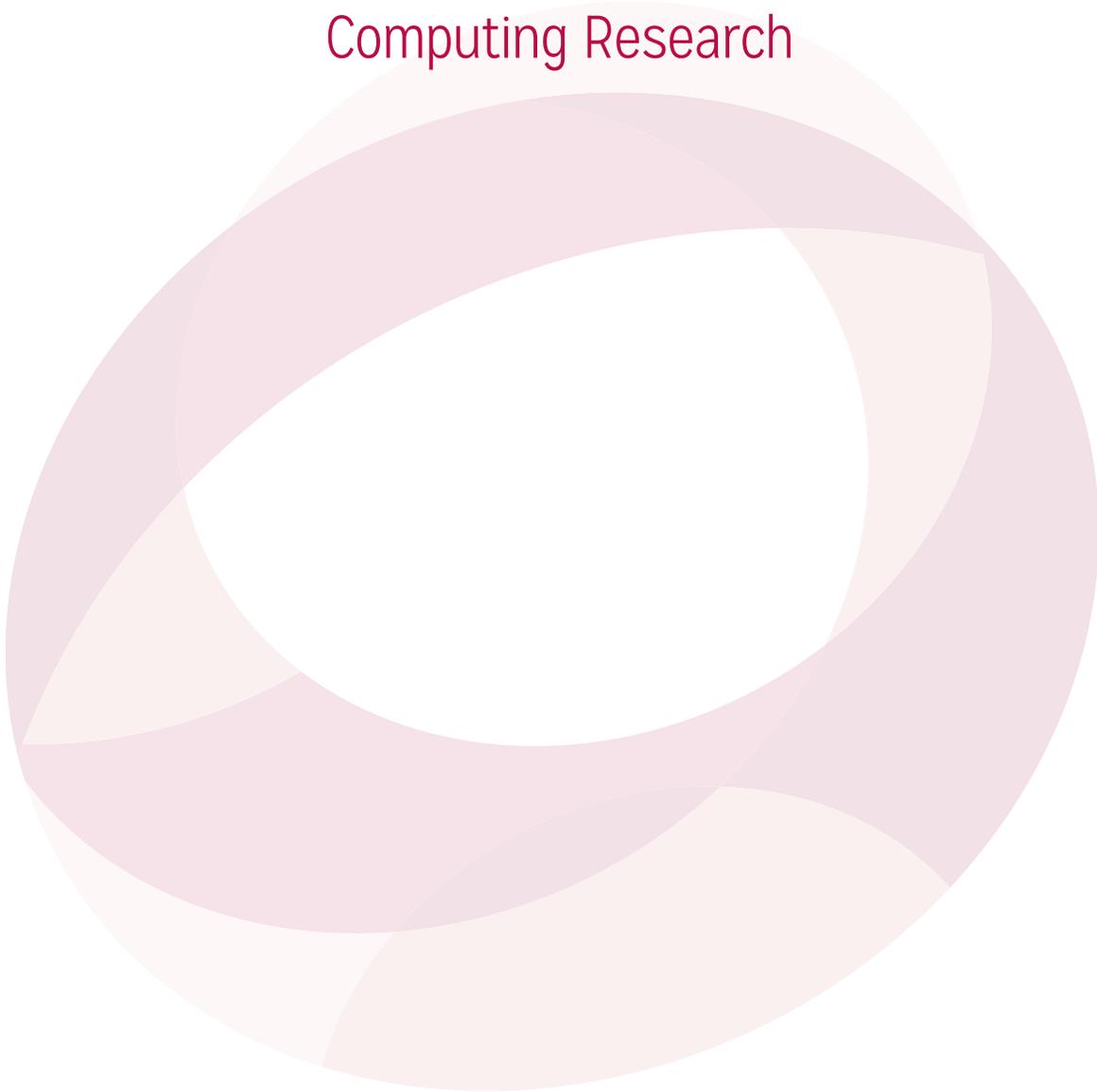

CCC Future of the Research Enterprise Task Force:

Benjamin Zorn (Microsoft Research), Tom Conte (Georgia Institute of Technology),

Keith Marzullo (University of Maryland), Suresh Venkatasubramanian (University of Utah)

This material is based upon work supported by the National Science Foundation under Grant No. 1734706. Any opinions, findings, and conclusions or recommendations expressed in this material are those of the authors and do not necessarily reflect the views of the National Science Foundation.

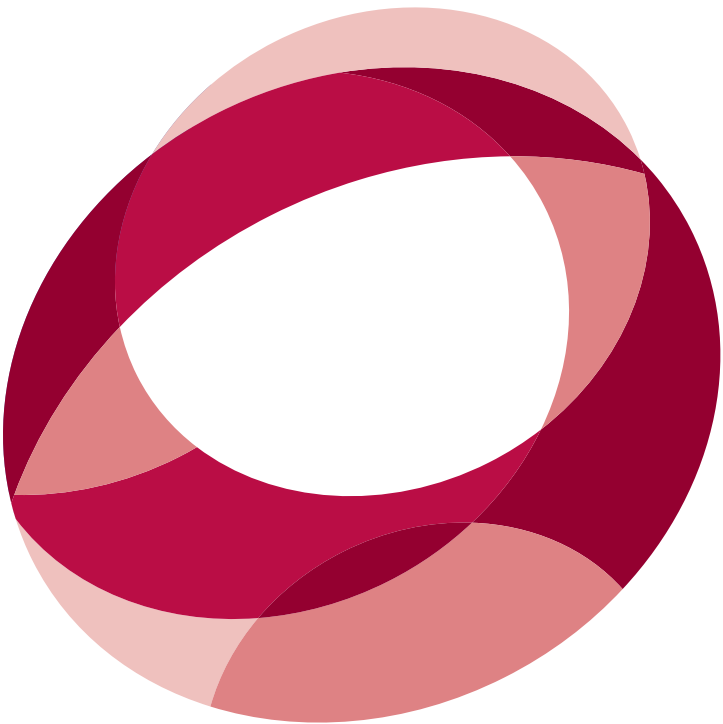

## Abstract

**Social and technical trends have significantly changed methods for evaluating and disseminating computing research.** Traditional venues for reviewing and publishing, such as conferences and journals, worked effectively in the past. Recently, trends have created new opportunities but also put new pressures on the process of review and dissemination. For example, many conferences have seen large increases in the number of submissions. Likewise, dissemination of research ideas has become dramatically easier for individuals even in the absence of peer review through publication venues such as arXiv.org and social media networks. While these trends predate COVID-19, the pandemic could accelerate longer term changes.

Based on interviews with leading academics in computing research (listed in the acknowledgements), our goals for this Computing Computing Consortium (CCC) task force white paper are to:

◗ **Present the trends observed.**

◗ **Discuss the impacts on the review and dissemination process.**

◗ **Suggest methods and recommendations to reduce the negative impacts of those trends.**

Our findings include:

◗ **Trends impacting computing research are largely positive** and have increased the participation, scope, accessibility, and speed of the research process.

◗ **Challenges remain in securing the integrity of the process**, including addressing ways to scale the review process, avoiding attempts to misinform or confuse the dissemination of results, and ensuring fairness and broad participation in the process itself.

Based on these findings, we recommend:

◗ **Regularly polling members of the computing research community,** including program and general conference chairs, journal editors, authors, reviewers, etc., to identify specific challenges they face to better understand these issues.

◗ **An influential body,** such as the Computing Research Association (CRA), **regularly issues a "State of the Computing Research Enterprise" report** to update the community on trends, both positive and negative, impacting the computing research enterprise.

◗ A deeper investigation, specifically to **better understand the influence that social media and preprint archives have on computing research,** is conducted.

◗ **Initiate an investigation of the impact of COVID-19** on the broader computing research enterprise, including the impact on evaluation and dissemination.

## Introduction

The process of conducting scientific research, specifically in terms of the review and dissemination of new ideas, has not changed dramatically in the last century. The main venues for evaluating and reporting new ideas continue to be conferences and journals and the process of evaluation continues to be through voluntary peer review, including program committees, editorial boards, etc. Computing research has followed this model with the significant change that many computing subfields consider publications in conferences as important or more important than publications in journals.[1] However, significant trends in computing research and the application of computing technology have resulted in pressure on the traditional review and dissemination methods from two sources: **pressure to scale** for reasons that include the widespread use of computing in society, and the influence of **new technology,** such as social networks that define new approaches to reviewing and dissemination. We consider each of these influences in more detail.

---

[1] As described in the CRA Best Practices memo: https://cra.org/resources/best-practice-memos/evaluating-computer-scientists-and-engineers-for-promotion-and-tenure/

**Trends in scale**

There are many ways to measure the **increasing impact that the field of computing has** had on society, including measures of enrollments in computer science programs,[2] increasing industry investment in key technologies such as artificial intelligence,[3] and the growth of the tech sector of the US economy.[4] Mirroring this growth, we see **submissions to top conferences across all of computing research growing**, especially in AI-related conferences.[5, 6] Another aspect of this trend is the **increase in papers appearing without review** in open document repositories, specifically arxiv.org where the number of AI papers appearing went from roughly 500 papers in 2010 to 13,000 papers in 2017.[7]

**Trends in efficiency**

Dramatic changes in technology have also created new capabilities to review and disseminate research ideas. In particular, the creation of social networks enables individuals to communicate directly with large numbers of like-minded colleagues, allowing ideas to be distributed without intermediation by authoritative bodies such as conference committees. Similarly, the cost of publication, which historically required the printing, shipping, etc., has reduced dramatically, enabling the cheap and rapid dissemination of ideas with fewer inefficiencies.

**Trends in analysis**

As the capabilities of computers increase, they can increasingly be used as part of the review and dissemination process. Computers are already routinely used to detect plagiarism, although such approaches can also be hacked. As AI capabilities, such as semantic word embeddings,[8] improve, computers are increasingly able to understand and analyze the content of publications for identifying related work, etc.[9, 10] For communication, machine-learning based recommendation systems are widely used in industry (e.g., for book and movie recommendations) and can be equally applied to help researchers understand what related research is most relevant.[11]

**Trends in participation**

Historically, contributions to computing research have occurred largely in papers written in English and presented at conference venues, often in the United States. Increasingly, major research contributions in computing research are occuring at non-US institutions. At the same time, machine translation between languages has improved dramatically, potentially reducing the language barrier in disseminating new ideas. Also, important research contributions in some areas, such as AI, come increasingly from industrial research efforts that

---

have the compute, data, and engineering resources to conduct experiments at much larger scale compared to their academic counterparts.

Based on our conversations with members of the computing research community, including recent program chairs of major conferences, we believe that these trends are having a significant impact, both positive and negative, on the process of reviewing and disseminating computing research. The research community should acknowledge these influences and take proactive measures to reduce the negative impacts of these changes.

## Impact on Review and Evaluation

### An Evolutionary Process

Our discussions highlighted a consensus that the **process of review in computing research has evolved rapidly over the last decade**. In addition, these **changes have been largely positive** and arise from an understanding of best practices combined with the availability of technology, such as conference review software, to easily apply them. Evolutionary practices that have become widely used include:

◗ Double blind reviewing,[12]

◗ Allowing author response to reviews,

◗ Creating independent external review committees to review submissions of members of the program committee,

◗ Adoption of a process for artifact evaluation and recognition,[13]

◗ Assigning shepherds to oversee paper revisions, and

◗ Conferences with rolling deadlines and multiple deadlines per year. VLDB was one of the early conferences that started this practice.

Some of these practices incorporate some of the strengths of traditional journal review, including some venues, such as HiPEAC and UIST, where the conference has become "journal-first", requiring journal acceptance before being presented in the conference. We also note that the degree to which these practices have been adopted varies across different sub-disciplines of computer science. As different communities (such as AI, CHI, databases, programming languages, etc.) develop new practices, the sharing of conference software between communities enables the transfer of practices effectively. Historically, this sharing is limited because different communities often use different reviewing software.

### Positive Revolutionary Changes

Another common theme we heard reflecting a positive impact on computing research is the degree to which **major technology shifts, including the Internet, cloud computing, and teleconferencing, have greatly enhanced the computing research process.**

These revolutionary changes have resulted in a **much broader global participation in the computing research process** and an **explosion of new research results**, especially in areas of intense commercial interest such as AI, computer vision, and natural language processing.

Many factors contribute to allowing greater participation in the computing research process including:

◗ The virtualization of major events, allowing remote participation. IEEE Collabratec is one example of an organizational effort to leverage technology for this purpose.[14]

◗ Low-cost, low-latency global access to research publications, documentation, and the researchers themselves.

---

[12] To see the status of double-blind reviewing in computing research conferences, visit https://double-blind.org/
[13] For a list of computing research conferences that review artifacts: http://evaluate.inf.usi.ch/artifacts
[14] https://ieee-collabratec.ieee.org/

- Shared implementations and data sets (via technology like github) and access to free compute resources via web-enabled infrastructures like Google's Colaboratory.[15]

- Virtualization container technology, such as Docker, allows entire computing environments for experiments to be archived and shared.

- Free high-quality training materials for implementation skills, basic technical background, and advanced computing research topics.

The COVID-19 epidemic dramatically highlights the degree to which the computing research process has evolved and impacted both the computing research and other scientific communities. Examples of this influence include:

- The broad adoption in different communities of preprint servers, such as arXiv, bioRxiv, and medRxiv has enabled a dramatic acceleration and scaling in the process of generating, sharing, and reviewing scientific research.

- By sharing both papers and data, other researchers can almost immediately check results. For example, a bioRxiv preprint paper that erroneously identified COVID-19 as human-made was determined to be erroneous by others within hours of its posting and removed the next day.[16] However, such an incident is anecdotal and does not necessarily counterbalance some of the negatives of preprint archives discussed below.

- Demonstrating the power of leveraging the cloud, the computing research community among many others has adapted its review process from almost entirely in-person review meetings and conferences to entirely virtual meetings. The pressure to convert events to be entirely virtual has enabled innovations that bring social opportunities previously only available to in-person participants to those attending remotely as well.[17]

## Negative Impacts on Review and Evaluation

While the effects of these trends on computing research have been largely positive, there are also side-effects with significant negative consequences. These negative effects fall into the following categories:

- **Strain on the review process due to scale** both in numbers of submissions and increasing diversity in research topics,

- **Incentives for unethical practices** due to the increasing commercial impact of computing technology, and

- **Increased pressure on authors to produce.**

**Many computing research conferences have seen dramatic increases in submissions** in recent years. CHI 2020, a top HCI conference, received 3,126 submissions that were overseen by 2 paper chairs, 38 subcommittee chairs, 467 associate chairs, and 3,072 reviewers.[18] Ultimately, 760 papers were accepted. Managing this complex and time-consuming process is a heroic effort by the organizers, especially when considering that almost all the work is voluntary. Technology for managing the submissions, committees, reviews, rebuttals, decisions, etc. has evolved over time but challenges remain. Clearly, for CHI and many other large conferences, **hierarchical decomposition (e.g., breaking the meeting into multiple, separately managed but coordinated sub-meetings) addresses some scaling problems.** Still, having so many submissions presents significant challenges to organizers whose job is to connect a large and diverse body of submissions with the appropriate subject-matter experts.

Conference chairs are **confronted with a greater diversity in the subject matter of submissions,** requiring them to identify and engage experts from research communities that might be quite different than their own. There are benefits to relatively small program committees including broader participation in discussions, greater social connection among the members, and the opportunity for mentoring among committee members. At scale such benefits are less likely and **the ability of different PC members to have awareness of many submissions or comment on them diminishes.**

---

[15] https://colab.research.google.com/notebooks/intro.ipynb
[16] https://www.statnews.com/2020/03/23/bioscience-publishing-reshaped-covid-19/
[17] https://iclr.cc/Conferences/2020/CallForSocials
[18] Katie Siek, personal communication. For more information about the CHI 2020 review process, see http://chipc.acm.org/2020/

The two factors of more diverse submissions and a reduction in the shared understanding of those submissions by the program committee make it more difficult for the committee to detect ethics violations. The kinds of unethical behaviors that have been observed include:

- **Collusion between PC members and authors,** which is especially possible when decisions are made virtually and program committee members can collude among themselves (e.g., with side-channel conversations) freely. For example, irregularities in the reviewing of an ISCA 2019 submission, which was connected with the tragic death of an author by suicide, have led to a joint IEEE TCCA and ACM SIGARCH investigation of the circumstances.[19]

- **Gaming the identification of author conflicts.** Authors are typically asked to self-report conflicts, and purposefully naming PC members that are potentially hostile as conflicts when they are not might lead to a more favorable review process.

- **Submitting papers to multiple venues simultaneously.** Plagiarism tools like TurnItIn[20] are increasingly used to detect overlap in paper submission and with previously published papers.

Other potentially negative outcomes due to scaling include:

- **Lack of vetting of PC members** due to the need for large committees that include experts from other research communities. Anecdotally, we have observed that for conferences with literally thousands of PC members, some PC members are identified only via an email address, which is easy to forge.

- **Conflict of interest challenges and committee size impacting the quality of review** due to lack of expertise. For example, if there are many subject matter experts on a PC, then getting external reviewers to review PC papers may be difficult due to lack of expertise.

- **Imbalance in numbers of junior and senior researchers** places additional review and mentoring burdens on senior members of the computing research community.

- **Higher stakes for authors.** Whether or not a paper is accepted at a major computing research conference can have significant consequences for the author's career. With increased competition, the pressure to succeed increases, increasing the potential for unethical behavior.

- **Increasing numbers of predatory journals and conferences.**[21] Due to the amount of competition in top conferences with exploding submission numbers, the "publish-or-perish" incentive leads authors to seek potentially predatory venues for their research.

- Broader concerns about the **relationship between national interests and the scientific process,** including computing research, have been raised[22]. For example, cases have been recorded of researchers not reporting affiliations with foreign military organizations.[23] Recently the National Science Foundation announced a new position, Chief of Research Security Strategy, specifically to consider these influences.[24]

---

[19] https://twitter.com/josep_torrellas/status/1158088204840591361?s=20

[20] https://www.turnitin.com/

[21] https://en.wikipedia.org/wiki/Predatory_publishing

[22] NSF Response to the JASON Report "Fundamental Research Security", https://nsf.gov/news/special_reports/jasonsecurity/NSF_response_JASON.pdf

[23] https://www.aspi.org.au/report/picking-flowers-making-honey

### External Forces

The COVID-19 crisis also highlights a sentiment heard in our conversations, that the **computing research process is already being influenced by significant external factors.** In particular, the unusual emphasis in computing research of publication in conferences and attending them to present important research has been disrupted by a global pandemic. Many other aspects of the computing research process, including education at all levels, research internships, and organizational meetings, have shifted to be entirely virtual. While the COVID-19 crisis will eventually end, the influence of this disruption is likely to last much longer. Recent conferences, such as ISCA 2020 and PLDI 2020, that were entirely virtual, reported record numbers of registrations.

And independent of COVID-19, the computing research community was already aware of the **climate impact of large amounts of travel that the research process encourages.** Organizations such as ACM SIGPLAN have started encouraging the systematic reporting of the carbon footprint of events and implementing mechanisms to ensure that events remain carbon-neutral by requiring the payment for carbon offsets.[25]

## Impact on Dissemination

While many scientific disciplines place greater emphasis on journals, computing research has emphasized the importance of conference publications over journals. Recently, a number of factors have both led to pressures on the traditional methods of research dissemination and technology (including conferences and journals), and created innovation opportunities for new methods.

### Shift to Preprint Archives

Pressures on traditional publication methods include:

- A **move to open access** research publications across numerous disciplines, reducing the financial incentives for the traditional publishers of such papers, including professional societies like IEEE and ACM as well as for-profit publishers such as Elsevier and Springer.

- The ready **accessibility of free preprints** of publications, reducing the value of providing access to research behind a pay-wall.

- The rise of **preprint archives,** such as arXiv, which provide both a centralized repository and expanded services around preprints, including services such as indexing, connections to social media, feedback mechanisms, etc.

- A **negative feedback loop** whereby universities, which are typically the major source of revenue for traditional publishers, choose not to renew their subscriptions due to budget pressures and the reasons above.

- **Unknown effects from this shift on professional societies.** Organizations such as the ACM and IEEE, which depend in part on membership and paid access to digital media for financial viability, are seeing decreases in membership and potential loss of revenue from paid access to digital media. There is also a "greying" of membership phenomenon wherein younger members in Computing see less benefit in being a member of a professional society.

- **Pressure from external parties, such as journalists, to report on the most recent advances** as soon as they are available. Given that journalists can greatly expand the visibility of new results, their decision to report on preprints impacts both the academic community and society at large.

---

[24] https://www.nsf.gov/news/news_summ.jsp?cntn_id=300086
[25] https://blog.sigplan.org/2019/07/17/acm-conferences-and-the-cost-of-carbon/

## Positive Effects from New Methods of Dissemination

As a result of these pressures, the financial viability of traditional publishers remains in question while the use of alternative publication methods, such as preprint archives, has grown dramatically. There are many positive impacts of this trend, some of which have been already mentioned, including:

- **Global availability of timely results,**
- **Reduction in barriers to entry** to less advantaged participants,
- **Integration** of documents, data, and software tools combined with added services, and
- Benefits of the **network effect,** where contributions and improvements to the shared preprint archive benefit all members of the community.

Similarly, the rapid dissemination and large audiences available for **sharing ideas on social media** has greatly increased conversations and sharing of research results on platforms including Twitter and Facebook. Many COVID-19 related research results, such as the latest results in genome sequencing from the Nextstrain project,[26] are posted multiple times per day via their Twitter feed. Individuals using Twitter benefit from the small size of posts, allowing them to **quickly process inputs from diverse sources on cross-cutting topics** that they might not otherwise have the time to understand. The visibility of relevant research results to the general public also creates a stronger connection with the research community.

## Negative Impacts from Dissemination via Preprint Archives and Social Media

The agile process of posting research results to preprint archives and social media **lacks the important element of review by subject matter experts.** There are numerous potentially negative consequences of this failing:

- **Incorrect ideas have the same status as well-researched ideas.**
- **Individuals can be misled** because they lack the skills to distinguish information from misinformation. Both understanding the methods of misinformation, and approaches to preventing it, have greatly increased in recent years for this reason.
- Proxies for authority, such as number of followers or social status, **can give individuals or organizations without expertise undue influence.** Conversely, the purpose of double-blind reviewing, which is widely believed to be an effective practice, is to avoid the authority of the author or the institution influencing a reviewer's decision to publish. Further, individuals or organizations with substantial resources and public relations expertise can use those resources, potentially, to promote their research results whether or not they have merit.
- The **amount of information available to an individual can overwhelm** their ability to process it.
- Some media, such as tweets, can be deleted, **encouraging the creation of dubious content.**

Many of these limitations are widely known and are active areas of computing research investment, including in areas of misinformation, bias and fairness, and creating reputation systems. In the next section, we outline areas of investment that will likely mitigate some of the greatest negative impacts we have discussed.

## Reducing Negative Impacts

We partition this discussion into impacts on the review and evaluation process and impacts on the dissemination process.

### Reducing Negative Review and Evaluation Impacts

Because the review and evaluation process for computing research is already largely mediated by software frameworks (for conference and journal review), solutions in this space that can be achieved through augmenting such frameworks are attractive. These include:

- Better support for automating the process of determining author conflicts. Research on this topic is already underway.[27]

- Greater use of tools to detect simultaneous submissions, submission overlap, and plagiarism.

- More consistent use of strong authentication mechanisms for committee members and reviewers.

- Greater awareness for authors of the existence of predatory journals and conferences, and encouragement of authors to avoid engaging with predatory venues.

---

[26] https://nextstrain.org/

To address issues related to the size of conferences and the climate impact of travel:

- Increase the number and prestige of regional computing research meetings as compared to annual global meetings. For example, instead of one ISCA,[28] there would be ISCA North America, ISCA Europe, etc. These regional meetings could still allow global participation via teleconferencing but could be scaled to have fewer submissions and smaller committees.

- Create "large-conference best practices and tool support" documents based on existing experiences with scale in conferences such as CHI.

- Better understand the pressures on authors and provide greater support within the community to address unnecessary pressure to generate results and publish.

### Reducing Negative Dissemination Impacts

At the heart of addressing the negative impacts of social media and preprint archive posting of computing research results is the need to vet material quickly and effectively by subject-matter experts. Here are a few suggestions for approaches in this direction:

- Encourage preprint archives to hide author name/affiliation until a certain level of vetting is accomplished (such as acceptance in a peer-reviewed venue), incentivizing the authors to obtain such vetting quickly.

- Conference steering committees could encourage authors to reference published and peer-reviewed prior work, when available, instead of referencing preprints.

- Automatically check submissions to preprint archives for baseline quality metrics, like potential plagiarism, relative completeness of citations to related work, practices of overuse of self-references, etc.

- While social media networks try to vet content for misinformation, perhaps academic bodies could create "research social media vetting services" that could be used to check accuracy of content in social media making claims related to computing research results.

- As journalists are increasingly a part of the research dissemination ecosystem, the computing research community should actively engage with journalists regarding the best practices around reporting new results.

### Conclusions

We have explored the impact of significant social and technology trends on the process of computing research review, evaluation, and dissemination. We conclude that **the impact of these trends has been enormous and mostly positive,** greatly enhancing the ability of individuals around the world to contribute rapidly and effectively to the body of computing research literature. **We have also identified significant challenges that the computing research community faces related to the negative impacts of these trends.** These challenges include:

- Pressure on the review and evaluation process due to the increased number and diversity of submissions, especially in subject areas with significant growth,

---

[27] "Pistis: A conflict of interest declaration and detection system for peer review management", ICMD 2019. https://dl.acm.org/doi/abs/10.1145/3183713.3193552

[28] ACM/IEEE International Symposium on Computer Architecture

- Pressure on authors to publish and compete in the face of greater competition,
- Increased incentives for unethical behaviors related to increased publication pressure and the difficulty of scaling the review process, and
- Ensuring that computing research results published on social media and in preprint archives has a sufficient level of vetting by subject matter experts.

### Recommendations

In light of these challenges, we suggest that an authoritative body in the computing research community, such as the CRA, institute the following activities related to monitoring the computing research process and reducing the negative impacts of the trends:

- **Broadly poll members of the computing research community,** including those who are involved in the review and evaluation process such as program chairs and general chairs, to better understand the negative impacts they are facing. Including all constituents, such as students, research faculty and industrial participants will provide a more balanced view of how these trends affect the entire community.
- **Encourage the integration of tools that support best practices,** such as double-blind reviewing, and new tooling to address emerging challenges, in widespread use throughout the community.
- **Regularly publish a "State of the Computing Research Enterprise"** report capturing the feedback from the community for the purpose of sharing an understanding of the challenges, tools, and best practices that emerge. Among other aspects, such a report may also include measures of demographics of authors (e.g., lay collaborator, interdisciplinary colleague, gender identity, race, "rank" – undergrad, PhD student, post doc, PI – industry versus academia, etc.) to better understand the breadth, inclusiveness and diversity of participation in computing research.[29]
- **Better understand the influence that social media and preprint archives have** on the review and dissemination of both important computing research and misinformation about computing research. Engaging with and understanding the role of journalists in the dissemination of computing research is also valuable.
- **Initiate an investigation of the impact of COVID-19 on the broader computing research enterprise,** including the impact on evaluation and dissemination. While there will be many studies of the impact of COVID-19, we believe that the computing research impact is sufficiently unique and important that it warrants a focused investigation. As we have mentioned throughout this report, COVID-19 has already had a significant impact on the computing research process. Looking forward, possible longer term impacts include changes in faculty hiring, research budgets, collaboration models, etc. Understanding the long-term implications of COVID-19 is a necessary step to anticipating the changes and adapting to them.

### Future Considerations

We have considered numerous aspects of the evolution of evaluation and dissemination of computing research but, to achieve succinctness, we have not considered other important aspects. These aspects include: a more systematic treatment of including input and output data sets as part of publication, the method and requirements for sharing code artifacts, methods around ensuring reproducibility, and accurately citing code and data sources in publications. Likewise, while we have enumerated some approaches taken by different conferences to incorporate aspects of journals, we have not advocated that specific practices should be more widely followed. We consider such topics important and hope that initiating a process to regularly survey such issues will bring them under consideration in the future.

---

[29] For example, early data suggests a negative impact of COVID-19 on women publishing academic papers: https://www.insidehighered.com/news/2020/04/21/early-journal-submission-data-suggest-covid-19-tanking-womens-research-productivity

## Acknowledgments

We received valuable input from numerous individuals throughout the course of this study. We gratefully acknowledge their input here:

- Andy Bernat (CRA),
- Emery Berger (University of Massachusetts, Amherst),
- Pernille Bjørn (University of Copenhagen, Denmark),
- Steve Blackburn (Australian National University),
- Khari Douglas (CCC),
- Ann Drobnis (CCC),
- Kathleen Fisher (Tufts University),
- Stephanie Forrest (Arizona State University),
- Ian Foster (Argonne National Laboratory and University of Chicago),
- Michael Franklin (University of Chicago),
- Juliana Freire (NYU),
- Jonathan Grudin (Microsoft),
- Kim Hazelwood (Facebook),
- Mark D. Hill (University of Wisconsin-Madison),
- Ran Libeskind-Hadas (Harvey Mudd College),
- Daniel Lopresti (Lehigh University),
- Melanie Mitchell (Portland State University),
- Rachel Pottinger (University of British Columbia),
- Jennifer Rexford (Princeton),
- Katie Siek (Indiana University),
- Shengdong Zhao (National University of Singapore),
- Ellen Zegura (Georgia Institute of Technology),

and members of the CCC Council and the CRA Board.